\documentclass[preprint,12pt]{elsarticle}

\usepackage{amssymb}

\journal{Annals of Physics}

\begin{document}

\begin{frontmatter}



\title{Thermal radiation and entanglement\\ in proton-proton collisions at the LHC}


\author[yup]{O.K. Baker}
\ead{oliver.baker@yale.edu}
\author[sbu,bnl]{D.E. Kharzeev\corref{cor1}}
\cortext[cor1]{corresponding author}
\ead{dmitri.kharzeev@stonybrook.edu}

\address[yup]{Department of Physics, Yale University, New Haven, CT 06520, USA}
\address[sbu]{Department of Physics and Astronomy, Stony Brook University, NY 11794, USA}
\address[bnl]{Department of Physics and RIKEN-BNL Research Center, \\ Brookhaven National 
Laboratory, Upton, NY11973, USA}

\begin{abstract}
The origin of the apparent thermalization in high-energy collisions is investigated 
using the data of the ATLAS and CMS Collaborations at the LHC. For this purpose, we analyze 
the transverse momentum distributions in the following proton-proton collision 
processes, all at $\sqrt{s} = 13$ TeV: {\it i)} inclusive inelastic $pp$ collisions; 
{\it ii)} single- and double-diffractive Drell-Yan production 
$pp \to \mu^+ \mu^- X$; and {\it iii)} Higgs boson production. We confirm the 
relation between the effective temperature and the hard scattering scale observed 
at lower energies, and find that it extends even to the Higgs boson production
process. In addition we find that the thermal component disappears in diffractive 
events (even though many charged hadrons are still produced). We discuss the implications of our study for the mechanism of multi-particle 
production -- in particular, we test the hypothesis about the link between quantum 
entanglement and thermalization in high-energy collisions.
\end{abstract}

\begin{keyword}
entanglement entropy \sep thermal radiation \sep particle physics 
\PACS 03.65.Ud \sep 13.85.Ni \sep 12.40.Ee

\end{keyword}

\end{frontmatter}


\section{Introduction}
\label{intro}
Understanding the dynamics of multi-particle production in high energy collisions 
remains a big challenge for theory. This is because the description 
of real-time evolution of a strongly coupled non-Abelian gauge theory is 
notoriously difficult. Nevertheless, the availability of the diverse high quality 
data on multi-particle production from the experiments at the LHC and RHIC should   
allow to motivate and inform the theory. In this work, we address the origin of 
the apparent thermalization in high-energy collisions that is usually inferred from 
the presence of the exponential component in the transverse momentum distributions 
of produced particles and the thermal abundances of the hadron yields (see \cite{Becattini:2009sc} for a review). The emergence 
of the thermal features in a high energy proton-proton collision is surprising, as 
the number of secondary interactions in this process is relatively low and does not 
favor thermalization through conventional final-state interaction mechanisms.

In this paper we will investigate the possibility that the apparent thermalization 
in high energy collisions is achieved during the rapid ``quench'' induced by the 
collision due to the high degree of entanglement inside the wave functions of the 
colliding protons \citep{kharzeev2017deep}. An example of such a ``quantum thermalization 
through entanglement'' is a recent experimental study of a quench in the Bose-Einstein 
condensate of Rb atoms where the entanglement was found to induce a rapid eigenstate thermalization 
\citep{kaufman2016quantum}.  Theoretical studies of quenches in entangled quantum systems 
described by $(1+1)$-dimensional conformal field theories 
\citep{calabrese2005evolution,calabrese2016quantum} indicate that at late times the system 
can be described by a generalized thermal Gibbs ensemble with an effective temperature 
set by the energy cutoff for the ultraviolet modes. 

Since a high-energy collision can be viewed as a rapid quench of the entangled partonic 
state \citep{kharzeev2017deep}, it is thus possible that the effective temperature inferred 
from the transverse momentum distributions of the secondaries in a collision can depend upon 
the momentum transfer, that is an ultraviolet cutoff on the quantum modes resolved by the 
collision. In analyzing the high-energy collisions with different characteristic momentum 
transfer $Q$ we thus expect to find different effective temperatures $T \sim Q$. We can also
look at the inelastic events characterized by a rapidity gap, where the proton is probed 
as a whole, and no entanglement entropy arises \citep{kharzeev2017deep} -- in this case, if the quantum entanglement 
is responsible for the thermalization, we expect no thermal radiation.  

In fact, it has been observed \citep{Bylinkin-Rostovtsev:2014} in deep-inelastic scattering 
at HERA that while the thermal component of hadron spectra is a prominent feature 
of inclusive events, this thermal radiation disappears in the events characterized by 
the rapidity gap. Since diffractive processes with a rapidity gap involve the entire wave 
function of the proton, there is no associated entanglement entropy  -- so this 
observation hints at a link between the entanglement and thermalization. The 
relation  between the effective temperature and the saturation momentum (that 
is an UV cutoff on the gluon modes in an inclusive interaction) that has been deduced 
\citep{thermal-dmitri:2014} from the inclusive data on inelastic collisions at RHIC 
energies also agrees with this hypothesis.  An alternative view is that the thermal 
radiation possesses a  
universal effective temperature $T \sim \Lambda$, where $\Lambda$ is the 
QCD scale that determines the mass gap in this theory. 

The large amount of data accumulated by the LHC experiments should 
allow to disentangle these distinct possibilities, and we will attempt to do 
it in this paper.  Specifically, we perform the comparison of the transverse 
momentum distributions in the following proton-proton collision processes at 
$\sqrt{s} = 13$ TeV: i) inclusive inelastic $pp$ collisions; 
ii) single- and double-diffractive Drell-Yan production $pp \to \mu^+ \mu^- X$; and iii) 
Higgs boson production. 
\vskip0.3cm

The paper is organized as follows. In section \ref{theory} we briefly 
summarize the theoretical ideas on the role of entanglement in high 
energy collisions, and discuss the possible link between entanglement 
and thermalization. In section \ref{mult-data} we analyze the transverse 
momentum distributions of charged hadrons produced in inelastic $pp$ 
collisions at $\sqrt{s} = 13$ TeV. Here we find both the ``thermal'' (falling off exponentially)  
and ``hard'' (falling off as a power) components, with the effective temperature and semi-hard 
scattering scales related in a way similar to what has been found at 
lower energies \cite{thermal-dmitri:2014}. In section \ref{diff-data} we 
analyze the single- and double-diffractive Drell-Yan production 
$pp \to \mu^+ \mu^- X$ at $\sqrt{s} = 13$ TeV. 
The dominant mechanism of $\mu^+ \mu^-$ production is the photon-photon 
fusion $\gamma \gamma \to \mu^+ \mu^-$, so these diffractive processes 
allow an analysis of the fragmentation of a high energy proton in an intense 
electromagnetic field produced by the other proton. We observe that the
thermal component of the hadron spectrum disappears in this class of 
events even though the events are inelastic and do produce many hadrons. In section \ref{higgs} we analyze the transverse momentum 
distribution of the Higgs bosons produced in $pp$ collisions at $\sqrt{s} = 13$ 
TeV. Surprisingly, we find that this transverse momentum distribution is also 
accurately described by the superposition of a power-like ``hard'' component 
with the hardness scale set by the Higgs boson mass $M_H$ and the thermal 
component with a very high effective temperature $T \sim M_H$, where the 
proportionality coefficient is close to the one observed for other inelastic 
processes. Finally, in section \ref{discussion} we discuss our results and 
their implications for understanding the mechanism of apparent thermalization 
in high energy collisions.

\section{Entanglement and thermalization in high energy collisions}\label{theory}

 Recently, it has been proposed that quantum entanglement is at the origin 
 of parton distributions measured in hard processes \cite{kharzeev2017deep}. 
 Let us briefly summarize these arguments here.  A hard process probes only 
 the part of the proton wave function that is localized in a region of 
 space that we denote $A$. For a hard process with a momentum transfer $q^2 = - Q^2$ 
 and Bjorken variable $x$, this region has a transverse size $\sim 1/Q$ and, 
 in the proton's rest frame, longitudinal size $\sim (mx)^{-1}$, where $m$ is 
 the proton mass. 

Let us denote by $B$ the region of space complementary to $A$, so that 
the entire space is $A \cup B$. The physical states inside the region $A$ 
probed by the hard process are states in a Hilbert space $\mathcal{H}_A$ 
of dimension $n_A$, and unobserved states in the region $B$ belong to 
the Hilbert space $\mathcal{H}_B$ of dimension $n_B$. The composite 
system in $A \cup B$ (the entire proton) is then described by the vector 
$|\Psi_{AB}\rangle$ in the space $\mathcal{H}_A \otimes \mathcal{H}_B$ 
that is a tensor product of the two spaces:
\begin{equation}\label{vecspace}
|\Psi_{AB}\rangle = \sum_{i,j} c_{ij}\ |\varphi_i^A\rangle \otimes |\varphi_j^B\rangle ,
\end{equation}
where $c_{ij}$ are the elements of the matrix $C$ that has a dimension 
$n_A \times n_B$. If one can find such states $|\varphi^A\rangle$ and 
$|\varphi^B\rangle$ that $|\Psi_{AB}\rangle = |\varphi^A\rangle \otimes |\varphi^B\rangle$, 
i.e. that the sum (\ref{vecspace}) contains only one term, then the state 
$|\Psi_{AB}\rangle$ is separable, or a product state. Otherwise the state 
$|\Psi_{AB}\rangle$ is entangled.

The Schmidt decomposition theorem states that the pure wave function  
$|\Psi_{AB} \rangle$ of our bi-partite system can be expanded as a 
{\it single} sum 
\begin{equation}\label{schmidt}
|\Psi_{AB} \rangle = \sum_n \alpha_n |\Psi_{n}^A \rangle |\Psi_{n}^B \rangle 
\end{equation}
for a suitably chosen orthonormal sets of states $|\Psi_n^A \rangle$ and 
$|\Psi_n^B \rangle$ localized in the domains $A$ and $B$, respectively, 
where $\alpha_n$ are positive and real numbers that are the square roots 
of the eigenvalues of matrix $C C^\dagger$.  In the parton model, we assume 
that this full orthonormal set of states is 
given by the Fock states with different numbers $n$ of partons.

 The density matrix of the mixed state probed in region $A$ can now 
 be written down as  
\begin{equation}
\rho_A = \rm{tr}_B\ \rho_{AB} = \sum_n \alpha_n^2\ |\Psi_n^A \rangle \langle \Psi_n^A |, 
\end{equation}
where $\alpha_n^2 \equiv p_n$ is the probability of a state with $n$ partons. 
The identification of the basis $|\Psi_n^A \rangle$ in the Schmidt decomposition 
(\ref{schmidt}) with the states with a fixed number $n$ of partons is natural -- only 
in this case we do not have to deal with quantum interference between states 
with different numbers of partons, as such interference is absent in the parton 
model. Because the parton model represents a description of QCD that is a 
relativistic field theory, the number of terms in the sum (\ref{schmidt}) (the 
Schmidt rank) is in general infinite. Note that a pure product state with no 
entanglement would have a Schmidt rank one.

The von Neumann entropy of this state is given by 
\begin{equation}\label{EE1}
S = - \sum_n\ p_n\ \ln p_n .
\end{equation}
This entropy results from the entanglement between the regions $A$ and $B$, and 
can thus be interpreted as the entanglement entropy. In terms of information 
theory, Eq. (\ref{EE1}) represents the Shannon entropy for the probability 
distribution $(p_1, ..., p_N)$. 
The QCD evolution equations can be used to evaluate the probabilities 
$p_n$, and thus the entanglement entropy (\ref{EE1}). 

After the hard scattering takes place, the mixed quantum state characterized 
by the entanglement entropy  (\ref{EE1}) undergoes the evolution towards 
the final asymptotic state of hadrons measured in the detectors. This final 
state is characterized by the Boltzmann entropy; how does this entropy relate 
to the initial entanglement entropy of the system? Does the produced Boltzmann 
entropy correspond to an entropy of a thermal ensemble?

To address these questions, let us consider the proton-proton collision in the reference frame where 
one of the protons is at rest. As discussed above, in this frame the partonic 
configuration of the high-momentum proton is prepared long before the 
collision, at a distance $\sim (m x)^{-1}$. The proton itself is an eigenstate 
$|\psi_0 \rangle$ of the QCD Hamiltonian $H_0$. When the collision takes 
place, this configuration undergoes a rapid ``quench'', and evolves according 
to a new Hamiltonian $H= H_0 + V(t)$ where $V(t)$ is the term induced 
by the inelastic interaction. Since an inelastic interaction in QCD is induced 
by the gluon exchange, the term $V(t)$ represents an effect of the pulse of 
the color field.  The onset of this pulse in a hard scattering with a hardness 
scale $Q$, by the uncertainty principle, is $\tau \sim 1/Q$ (we write it in the 
comoving frame, so $\tau$ is the proper time). Since this time is short on the 
QCD scale, $\tau \ll 1/\Lambda$, the quench creates a highly excited multi-particle state.

For the case of a short pulse of (chromo)electric field, the produced particles 
have thermal-like exponential spectra with an effective temperature of 
$T \simeq (2 \pi \tau)^{-1} \simeq Q/(2 \pi)$  \citep{kharzeev2005color}. 
The derivation in \citep{kharzeev2005color} involved a semiclassical 
approximation, but the same result holds for specific time profiles of the 
pulse when exact solutions can be found  \citep{dunne1998qed}. The 
thermal spectrum in this case can be attributed to the emergence of an 
event horizon formed due to the acceleration induced by the electric 
field \citep{kharzeev2005color,Castorina:2007eb}. 

These arguments point to the proportionality between the momentum scale 
$Q$ in an inelastic interaction and the effective temperature $T$ inferred 
from the transverse momentum distributions \citep{kharzeev2005color,kharzeev2007multiparticle,thermal-dmitri:2014}:
\begin{equation}\label{prop}
T = c\ \frac{Q}{2\pi}, 
\end{equation}
where $c$ is a universal (energy-independent) coefficient of order one. In an 
inclusive inelastic event, the scale $Q$ has to be identified 
\citep{kharzeev2005color,kharzeev2007multiparticle,thermal-dmitri:2014} with 
the ``saturation momentum'' $Q_s$ \cite{Gribov:1984tu,McLerran:1993ni,McLerran:1993ka} that depends on the Bjorken $x$ and thus 
on the energy of the collision and the rapidity at which the measurement of 
the spectra is performed. In a hard process,  the scale $Q$ is set by the 
kinematics of the process.

The emergence of thermal behavior in an entangled quantum system 
undergoing a quench has been recently observed in Bose-Einstein 
condensate of Rb atoms \citep{kaufman2016quantum}. The effective 
temperature was found to depend on the properties of the quench, similarly 
to the situation discussed above. 

Is it possible to predict the amount of produced 
Boltzmann entropy if one knows the initial entanglement entropy? 
In the case of a high energy collision, this would allow to predict the 
produced entropy if the parton distributions (interpreted in terms of 
entanglement entropy \citep{kharzeev2017deep}) are known. The comparison 
to the LHC data on hadron multiplicity distributions performed in 
\cite{kharzeev2017deep}  indicates that the produced Boltzmann entropy 
is quite close to the initial entanglement entropy. 

Unfortunately, very little is known at present on general grounds about the transformation 
of the entanglement entropy into the Boltzmann entropy following the 
quench. This problem is important and emerges in many areas of 
physics -- for example, solving it would enable understanding of qubit 
decoherence in quantum computers. The theoretical results available 
at present are mostly limited to the case of Conformal Field Theory (CFT). 
In particular, it is known \citep{calabrese2005evolution,calabrese2016quantum} 
that for a rapid quench (such as the one that occurs in a high-energy collision) 
in a $(1+1)$ dimensional CFT the entanglement entropy of a segment of length 
$L$ first grows linearly in time, until $t \simeq L/2$, and then saturates at the 
value 
\begin{equation}\label{quench}
S(t) \simeq \frac{c}{3} \ln \tau_0 + \frac{\pi c L}{12 \tau_0},
\end{equation}
where $c$ is the conformal charge of the CFT, and $\tau_0^{-1}$ is the energy 
cutoff for the ultraviolet modes. Comparing this to the entropy of a thermal 
$(1+1)$ dimensional system at a temperature $T$, $S_{therm} \simeq \frac{c}{3} L T$, 
we infer that the effective temperature is $T \sim \tau_0^{-1}$. 
Drawing an analogy to our case of a $(3+1)$ dimensional hard collision, we identify 
$Q = \tau_0^{-1}$, and expect to find an effective temperature $T \sim Q$, in 
accord with our previous arguments. 

\begin{figure}[ht!]
   \centering
   \begin{center}
   \vspace{-4cm}
 \hspace{-4cm}  \includegraphics[width=\columnwidth]{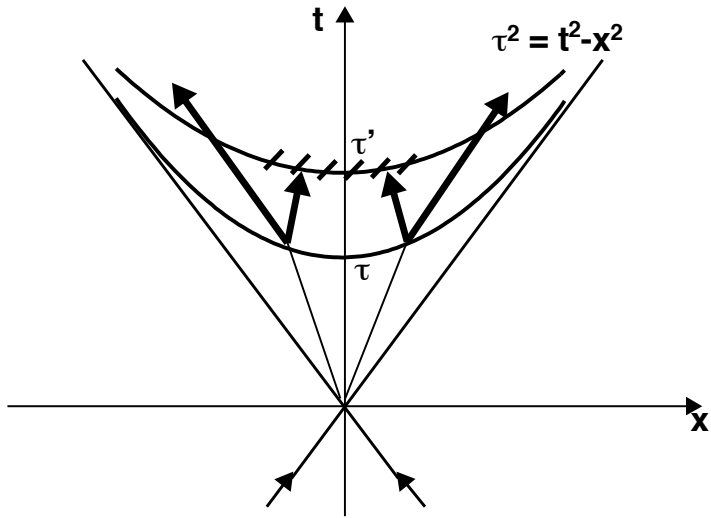} 
   \caption{A sketch illustrating quantum thermalization through entanglement in a high energy collision. The entangled particle pairs produced at a proper time $\tau$ contribute to the entanglement entropy in the rapidity interval shown by the dashed segment of the curve at a proper time $\tau' > \tau$.  
   }
   \label{fig:sketch}
   \end{center}
\end{figure}

The interpretation of the result (\ref{quench})  is the following \citep{calabrese2005evolution,calabrese2016quantum}. The quench leads to the production of entangled (quasi)particle pairs, since what used to be the ground state of the undisturbed Hamiltonian $H_0$ is a highly excited state of the Hamiltonian after the quench, $H = H_0 + V(t)$. The entangled pairs produced by the quench propagate along the light cone, and contribute to the entanglement entropy of the segment of length $L$ if only one particle of the pair is detected within this segment. Shortly after the quench, only particle pairs produced near the boundary of the segment thus contribute to the entanglement, and the entanglement entropy is not extensive in the length $L$. However, at times $t > L/2$, even in the center of the segment one can detect a particle whose entangled partner is outside of the segment -- this means that the entanglement entropy receives contributions from the entire segment, and should scale extensively in $L$ in accord with the result (\ref{quench}). This scaling is a necessary condition for an effective thermalization. For a quench induced by a high-energy collision, we sketch the resulting picture of thermalization from entanglement in Figure \ref{fig:sketch}. Note that the hardest quasiparticle modes that propagate along the light cone thermalize first. For the softer particles that propagate in the interior of the light cone, it takes a longer time to thermalize, i.e. to exhibit an extensive scaling of the entropy.
  
It is instructive to point out the difference in the mechanisms of thermalization expected at weak and strong coupling. At weak coupling, the ``bottom-up" thermalization mechanism \cite{Baier:2000sb}
also yields an effective temperature $T \sim Q_s$ in inelastic high energy collisions. However the thermalization in this picture begins from the soft, low-momentum modes that eventually draw the energy from the harder modes; the thermalization of the hard, high-momentum modes is thus expected to take a parametrically long time proportional to the inverse power of the (small) coupling constant \cite{Baier:2000sb}. On the other hand, in strongly coupled entangled systems the process of thermalization is fast and determined by the size of the system and the parameters of the quench; moreover, it starts from the {\it hardest} modes resolved in the process. In the dual holographic description of conformal field theory, this process is described by the formation of trapped surface near the Minkowski boundary that then falls into the AdS bulk, corresponding to the spreading of thermalization from hard to soft modes \cite{Lin:2006rf,Balasubramanian:2011ur}. A similar picture emerges from the analysis of entanglement entropy in an expanding string \cite{Berges:2017zws}, where the entropy has been found to have a thermal form with an effective temperature $T \sim 1/\tau$ at early time $\tau$.
\vskip0.3cm

The arguments presented above are qualitative at best, and can definitely 
be questioned. Nevertheless, we feel that they provide enough motivation 
to look into the structure of inelastic collision events at high energies, and 
to explore the possible relation between the effective temperature and the 
hard scale of the collision. We will now proceed to performing such an analysis.

\section{Charged hadron transverse momentum distribution}\label{mult-data}

Data from proton-proton collisions at 13 TeV center of mass energy yielding multiple
charged particles in the final state have been recorded by the ATLAS collaboration
at CERN's LHC
in 2016 \cite{ATLAS-pp01:2016}.  The dataset corresponds to an integrated luminosity 
of 151 $\mu {\rm b}^{-1}$ for charged particles that have transverse momenta greater 
than 100 MeV/c and absolute pseudorapidity  of less than 2.5.  Events that contain two 
or more charged particles in the final state were selected for analysis.  In order to remove
the presence of strangeness or heavier flavor charged particles from the sample,
final state hadrons that originate in the primary pp interaction and that have a 
lifetime of greater than 30 ps were excluded from the final selected events.  
\begin{figure}[ht!]
   \centering
   \begin{center}
   \includegraphics[width=\columnwidth]{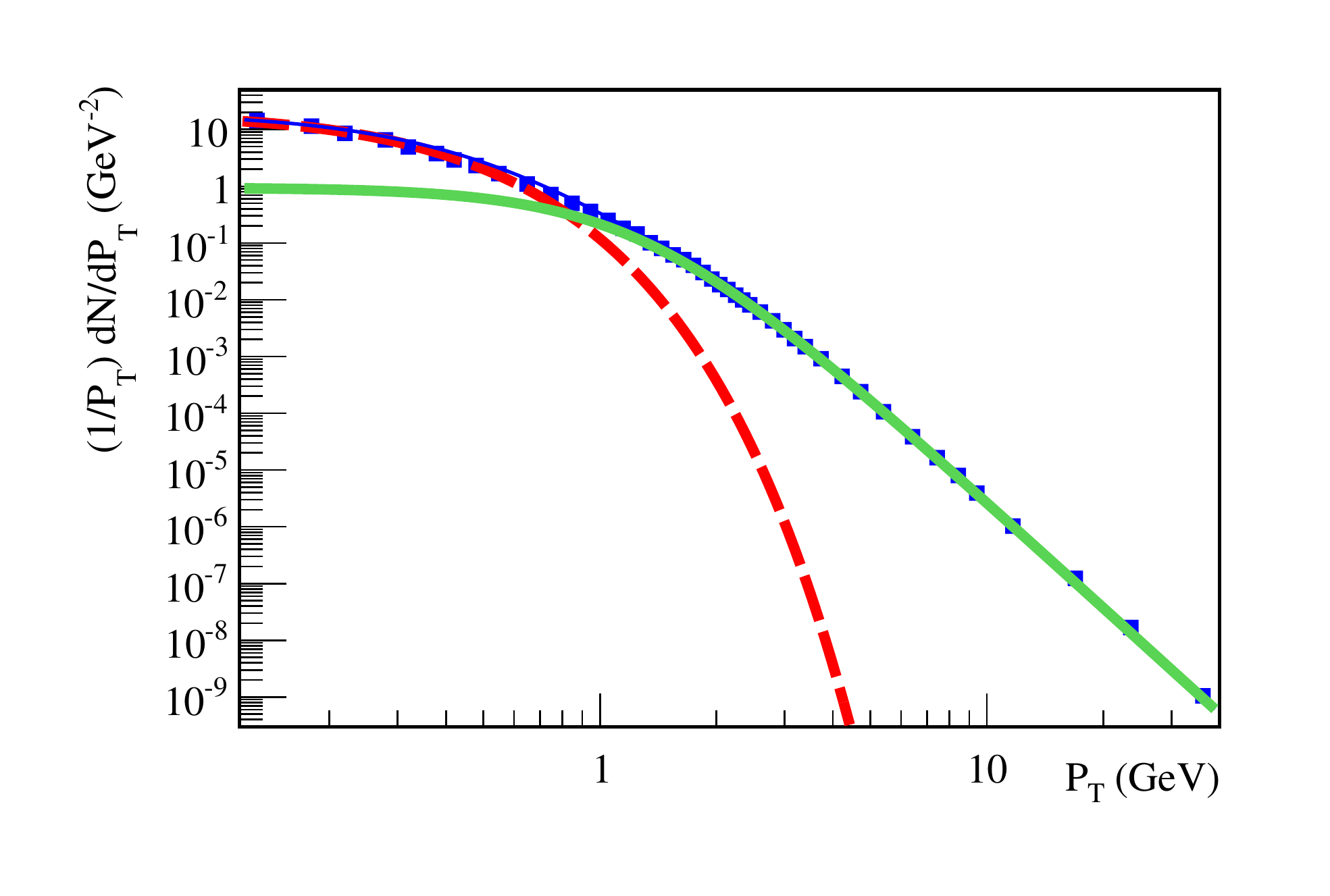} 
   \caption{Transverse momentum distribution of charged hadrons  in proton-proton collisions at $\sqrt{s} = 13$ TeV.
   The curves shown are exponential (red dashed) and power law (green solid) corresponding to the 
   thermal and hard scattering contributions respectively, and the sum of these two contributions (blue, thin solid).}
   \label{fig:proton-proton-pt}
   \end{center}
\end{figure}
Additionally, secondary charged particles
that are the result of particle decays from this dataset that have a lifetime greater than 
30 ps were also excluded.  The dataset effectively excluded charged strange baryons from 
the results used in the present analysis.  Comparisons with event generators Pythia
8  and EPOS (LHC tune) indicate that the
majority of selected events are non-diffractive and that the process is dominated
by t-channel gluon exchanges.  Transverse momentum bin widths from 0.1 GeV at low
$P_T$ to as large as 20 GeV at higher $P_T$ were used in the analysis \citep{ATLAS-pp01:2016}.

The normalized charged hadron transverse momentum distribution is 
shown in Figure~\ref{fig:proton-proton-pt}.
The thermal component is shown by the exponential, red
dashed curve; we parameterize it as 
\begin{equation}
{1\over N_{ev}} {1 \over 2\pi P_T}{d^2N_{ev} \over d\eta d P_T} \sim \rm A_{\it therm}exp(-m_T/T_{th}), 
\end{equation}
where the transverse mass $m_T$ is given by $m_T \equiv \sqrt{m^2 + P_T^2}$ ($m$ is the hadron mass; we assume that the spectrum is dominated by pions), and $T_{th}$ is an effective temperature.   The hard scattering (power law, green solid curve) component is parameterized as in  
\cite{thermal-dmitri:2014},
\begin{equation}
{1\over N_{ev}} {1 \over 2\pi P_T}{d^2N_{ev} \over d\eta d P_T} \sim \rm {A_{\it hard} \over \left(1+{m_T^2 \over T^2\cdot n}\right)^n},
\end{equation}
where $T$ and $n$ are parameters to be determined from the fit. 
 The sum of the two terms is shown by the blue solid curve.  
 
The value $\rm T_{\it th} = 0.17$ GeV describes well the experimental transverse momentum distribution; it agrees with that expected from the extrapolation of the relation \cite{thermal-dmitri:2014} deduced at lower energies
\begin{equation}\label{param_th}
\rm T_{\it th} = 0.098 \cdot \left(\sqrt{s/s_0}\right)^{0.06} \ {\rm GeV}
\end{equation}
to the LHC energy of $\sqrt{s} = 13$ TeV; $s_0 = 1\ {\rm GeV}^2$.  Similarly, the hard scale parameter 
$\rm T$ extracted in \cite{thermal-dmitri:2014} is 
\begin{equation}\label{hard_scale}
\rm T = 0.409 \cdot \left(\sqrt{s/s_0}\right)^{0.06} \ {\rm GeV}.
\end{equation}
Note that the parameterizations (\ref{param_th}) and (\ref{hard_scale}) imply that the effective temperature $\rm T_{\it th}$ is proportional to the hard scale $\rm T$, in accord with our discussion in section \ref{theory}.

Our fit to the charged hadron transverse momentum distribution yields the hard scale parameter $T = 0.72$ GeV and $n = 3.1$. 
This value of $T$ is in agreement with the extrapolation of (\ref{hard_scale}) to 
$\sqrt{s} = 13$ TeV, but the value of $n$ is smaller, reflecting the slower fall-off of the transverse momentum distribution at the LHC energy.

Let us define the ratio $R$ of the integral under the power law (hard scattering) curve and the 
integral under the total (hard scattering plus thermal component) curve of the fit in 
Figure~\ref{fig:proton-proton-pt}: 
\begin{equation}\label{ratio}
\rm R = {power \over power + exponential} .
\end{equation}
We find for it  the value of $R \simeq 0.16$, in agreement with the ratio calculated from  the charged hadron spectra in 
inelastic proton-proton collisions at ISR energies of $\sqrt{s} = 23, 31, 45, {\rm and}\ 53$ GeV \cite{Bylinkin-Rostovtsev:2014}.  

\section{Di-muon pair transverse momentum distribution\\ from $\gamma \gamma$ scattering in proton-proton collisions}\label{diff-data}

Proton-proton ($pp$) collisions at the LHC often proceed through the 
photon-photon ($\gamma \gamma$) interactions. In this case, the final state of the collision contains the protons, or the products $X', X''$ of their diffractive dissociation. 
The ATLAS collaboration made 
measurements of the reaction
\begin{equation}\label{gamgamtomumu}
pp (\gamma \gamma) \rightarrow \mu^+ \mu^- X' X''
\end{equation}
at 13 TeV center of mass energy in pp collisions \citep{ATLAS-mumu01:2017}.  The relevant Drell-Yan (DY) production 
processes are  
exclusive production (with two intact protons in the final state), single diffraction (in which one of the incident protons dissociates into an inelastic state), and double diffraction (in which both of the incident protons dissociate).
Selection of the exclusive $\gamma \gamma \rightarrow \mu^+ \mu^-$ process
was implemented by only including events that have both a $\mu^+$ and
$\mu^-$ track while excluding events that show additional charged particle
activity with transverse momenta greater than 400 MeV and within the pseudorapidity
range  considered here.  DY and multijet contributions, which are
backgrounds to the exclusive reaction, are vetoed with these cuts.  Additional
DY vetoing is achieved by excluding events that yield a di-muon invariant
mass greater than 70 GeV. The analysis performed in \citep{ATLAS-mumu01:2017} shows that at the transverse momenta of the Drell-Yan pair below 1.5 GeV the DY production is dominated by the exclusive process, whereas at larger transverse momenta the single and double diffractive processes with inelastic final states dominate.


In the most recent ATLAS analysis of the reaction (\ref{gamgamtomumu}) care was taken to 
select diffractive events that proceed through the $\gamma \gamma$ scattering.
  As argued in \citep{kharzeev2017deep,thermal-dmitri:2014}  such 
diffractive events 
are expected to have a suppressed thermal (exponential) component.  
This is because in these diffractive processes the photon interacts coherently with the entire proton, and no entanglement entropy arises, as discussed in section \ref{theory}. As the presence of the thermal component in this approach is the consequence of the entanglement, we expect it to be absent in diffractive events.

\begin{figure}[ht!]
   \centering
   \begin{center}
   \includegraphics[width=\columnwidth]{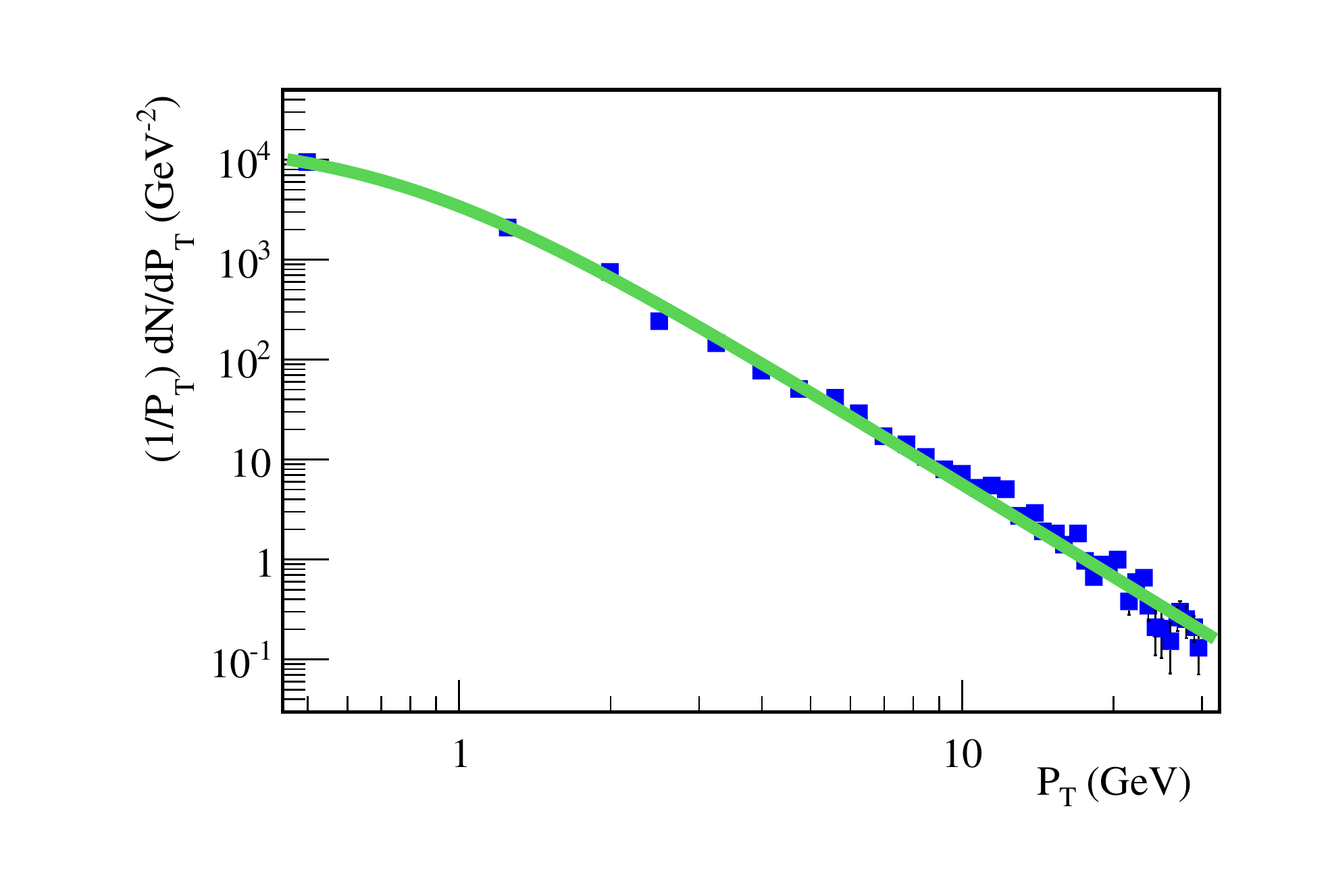} 
   \caption{Transverse momentum distribution of normalized event distribution 
     ${1\over P_T} {dN_{\mu \mu} \over dP_T}$  in units of GeV$^{-2}$ versus 
     the transverse momentum muon pair transverse momentum in units of GeV 
   for the (pp)($\gamma \gamma) \ \rightarrow \ \mu^+ \mu^- $(pp) reaction.  The
   curve shown (green, solid) is the power law contribution corresponding to the hard-scattering
   process.}
   \label{fig:muonmuon-pt}
   \end{center}
\end{figure}

Figure~\ref{fig:muonmuon-pt} shows the transverse momentum distribution in 
the case of $\gamma \gamma$ production of di-muon pairs in proton-proton
collisions at 13 TeV center of mass energy;   the transverse momentum
bin widths of 1.3 GeV were used in \citep{ATLAS-mumu01:2017}.  As can 
be seen from Fig.~\ref{fig:muonmuon-pt}, the hard
scattering term alone describes well the distribution, and there is no thermal
(exponential) component visible in the distribution.  
 The ratio $R$ defined in the
previous section in this case is $R \simeq 1$, in agreement with our theoretical expectations and the  previous data for 
$\gamma \gamma$ scattering at OPAL at $\sqrt{s} = 15\ \rm{and}\ 35$ GeV that also show no thermal component, with $R$ close to one.

\section{Higgs boson transverse momentum distribution}\label{higgs}

With the newest (and final) member of the Standard Model now discovered, we can 
 investigate whether the transverse momentum distribution of the Higgs boson is affected by the thermalization processes. While there is little doubt that the integrated cross sections of the Higgs production in general are adequately described by perturbation theory (see \cite{Dittmaier:2011ti} for a review), 
 it is possible that the QCD radiation in this process, and thus the Higgs boson transverse momentum distributions, are affected by the entanglement. 
 
  The Higgs boson transverse 
momentum distributions have been measured by both ATLAS and CMS collaborations
in the discovery mode channels: Higgs boson decays to four leptons (electrons and muons)
\citep{ATLAS-higgs4l:2017} and Higgs boson decays to $\gamma \gamma$
\citep{ATLAS-higgsgamgam:2017} \citep{CMS-higgsgamgam:2017}.   The data considered here are for
proton-proton collisions at $\sqrt{s} = 13$ TeV center of mass energy collected during Run 2 
in 2015 and 2016.

\subsection{Higgs boson decay to $\gamma \gamma$}
The dominant particle level process in the reaction $\rm pp \rightarrow H \rightarrow \gamma \gamma$
is gluon-gluon fusion, followed by the relatively less frequent vector boson fusion (VBF),  associated
production with top quarks (ttH) and associated production with a vector boson (VH).  The
fiducial cross sections for the reactions considered here are defined as the two photon final
states where the photons are well isolated and are restricted to the absolute pseudorapidity
region $\vert \eta \vert \le 2.37$, and where the leading and subleading photons satisfy
the requirement that the transverse momentum - diphoton invariant mass ratio
${P_T/m_{\gamma \gamma}}$ is greater than 0.35 and 0.25, respectively.    Photons must
have a transverse momentum greater than the threshold of 25 GeV, and only photons
that are detected outside of the ATLAS detector crack region, $1.37 \le \eta \le 1.52$ in pseudorapidity
are retained.   The diphoton invariant mass for Higgs boson
reconstruction is restricted to an invariant mass range between 105 GeV and 160 GeV,
inclusive. 

\begin{figure}[ht!]
   \centering
   \begin{center}
   \includegraphics[width=\columnwidth]{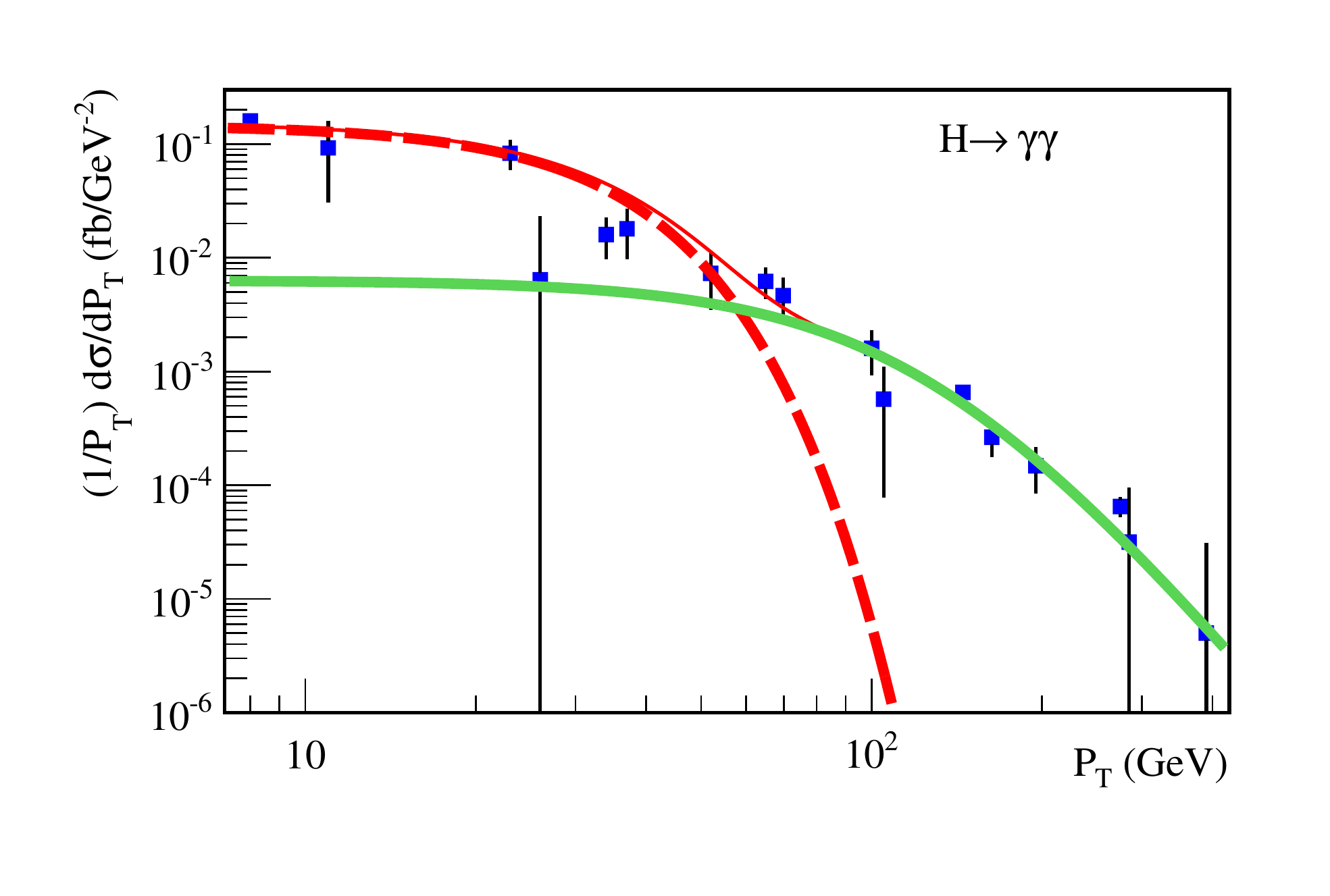} 
   \caption{Transverse momentum distribution of the Higgs bosons reconstructed from 
   the $H \rightarrow \gamma \gamma$ decay in proton-proton collisions at 
   $\sqrt{s} = 13$ TeV.  The curves shown are exponential (red dashed) and power law
   (green, solid) components corresponding to thermal and hard scattering contributions respectively; the
   sum of the two contributions is shown by blue, thin solid curve.}
   \label{fig:higgsgamgam}
   \end{center}
\end{figure}

Since the fiducial volumes of both ATLAS and CMS analyses are not too different
given the uncertainties in the measurements, the results from both experiments are included in
the current analysis.
In Figure~\ref{fig:higgsgamgam} the transverse momentum distribution of the Higgs bosons 
is shown in the range from 8 GeV to 390 GeV for combined ATLAS and CMS data.  As can be seen from Fig.~\ref{fig:higgsgamgam}, there clearly are both the hard scattering (power law) and thermal (exponential) components in the transverse momentum distribution, similarly to the case explored in section \ref{mult-data}. In fact, due to the much larger range of the available transverse momenta, the separation between the hard and thermal components is even more defined.

The power-law and exponential
distributions yield an effective temperature $T_{th} \simeq 3.5$ GeV and the hard scale parameter $T \simeq 14.4$ GeV that are about 20 times larger than the values derived from the charged hadron data in section \ref{mult-data}.  Interestingly, the ratio $R$ defined by (\ref{ratio}) and extracted from Figure~\ref{fig:higgsgamgam}  is $R = 0.15 \pm 0.05$ that is very close to the one determined from the charged hadron  
distribution in proton-proton collisions studied in section \ref{mult-data}, $R = 0.16 \pm 0.05$.

\subsection{Higgs boson decay to four leptons}

The Higgs boson decays to the additional high resolution final state channel 
(four leptons) were used to extract the transverse momentum ($P_T$) distributions
in both ATLAS \citep{ATLAS-higgsgamgam:2017} and CMS \citep{CMS-higgsgamgam:2017}.
\begin{figure}[ht!]
   \centering
   \begin{center}
   \includegraphics[width=\columnwidth]{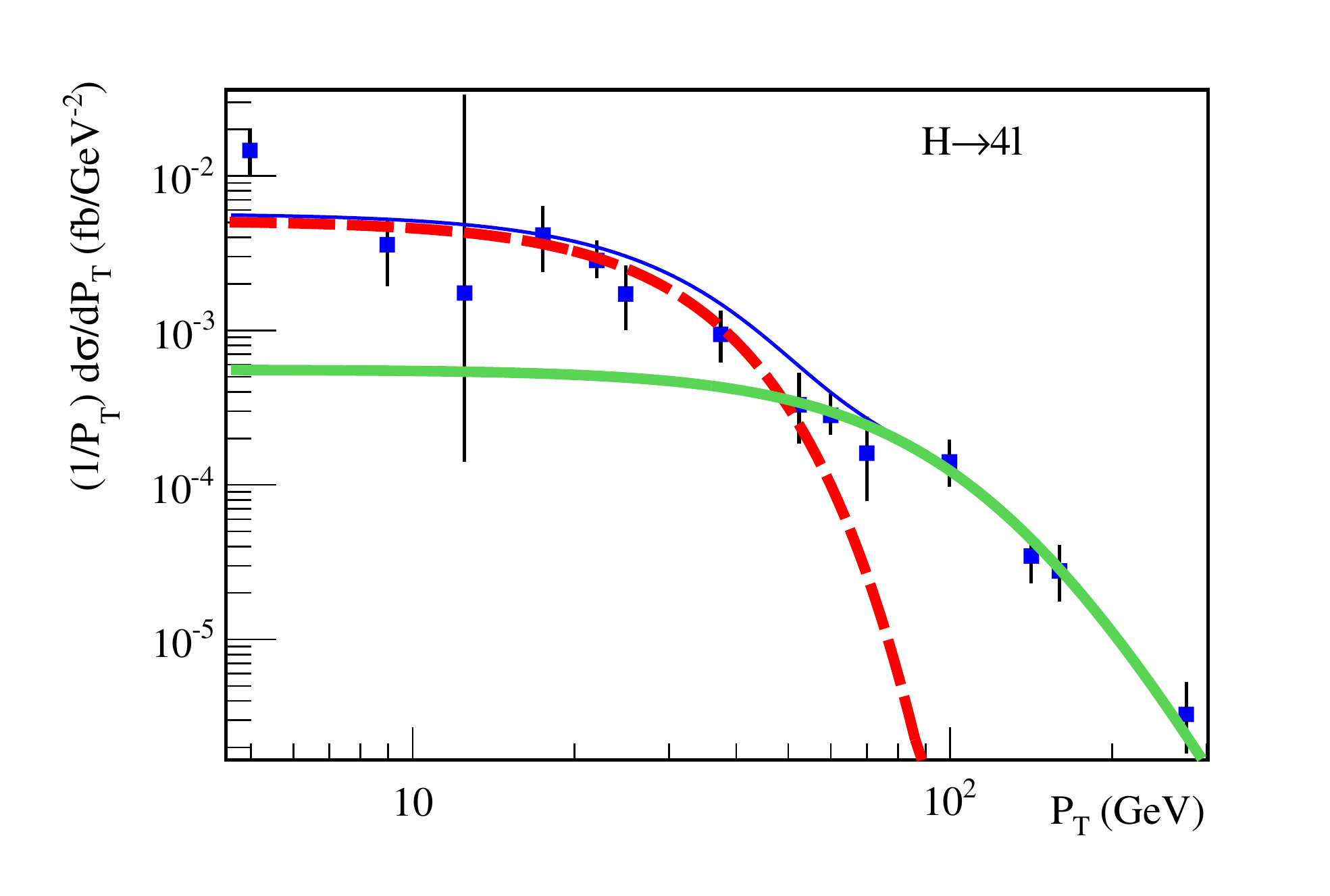} 
   \caption{The differential cross section of the Higgs boson production reconstructed from the $\rm H \rightarrow 4l $ (electrons, muons) decay in proton-proton collisions at 
   $\sqrt{s} = 13$ TeV.  The curves shown are defined as before.}
   \label{fig:higgsfourl}
   \end{center}
\end{figure}
Muon (electron) identification requirements are transverse momentum thresholds 
of 5 (7) GeV and an absolute pseudorapidity window of $\vert \eta \vert$ less 
than 2.7 (2.54) in ATLAS.  Systematic uncertainties on the detector correction and acceptance 
factors are at most 3.2\% and are mostly less than 1\%.  

Shown in Figure~\ref{fig:higgsfourl} is the normalized differential cross section,
${1 \over p^T} {d\sigma \over dp^T}$ in units of $\rm fb/GeV^2$ between 5 and 275
GeV transverse momenta for the $H \rightarrow ZZ^* \rightarrow 4l$ reaction.  Just
as for the case $H \rightarrow \gamma \gamma$ in the previous subsection, there
is here also a clear hard scattering component as well as a thermal component to
the full distribution.  (The curves are defined as before).  
As in the $H \rightarrow \gamma \gamma$ distribution described in
the previous subsection, the power-law and exponential components
yield an effective temperature and the hard scale parameter that are about 20 times larger than those determined from the charged hadron spectrum.
The ratio R calculated in the 4l case (see equation (\ref{ratio})) is $R = 0.23\pm 0.05$, which is consistent within the error bars with the value $R = 0.15 \pm 0.05$ extracted from the $H \to \gamma \gamma$ decay mode.
\vskip0.3cm

Table~\ref{temp} presents a compilation of the effective temperatures, hard scale parameters and the ratio R (defined by (\ref{ratio}) for the processes considered in this paper.

\begin{table}[htbp]
\caption{The effective temperature $T_{th}$, the hard scale parameter $T$, and the fraction of the hard component in the spectrum (\ref{ratio}) for different processes. 
}
\begin{center}
\begin{tabular}{cccc}
      $\rm T_{th}$, GeV  & T, GeV   & R & process\\
      \hline
     $0.17 \pm 0.03$  & $0.72 \pm 0.1$ & $0.16 \pm 0.05$  & pp $\rightarrow$ charged hadrons \\
     \hline
     none   & $0.1 \pm 0.02$ & $1.0 \pm 0.1 $ & pp $(\gamma \gamma) \rightarrow (\mu \mu)$pp   \\
     \hline
      $3.5 \pm 0.7$  & $14.4 \pm 0.3$  & $0.15 \pm 0.05$ & pp$\rightarrow {\rm H} \rightarrow \gamma \gamma$ \\
      \hline
      $3.5 \pm 0.7$  & $14.4 \pm 0.3$  & $0.23 \pm 0.05$ & pp$\rightarrow {\rm H} \rightarrow 4{\rm l} \ ({\rm e},\mu )$ \\
      \hline
\end{tabular}
\end{center}
\label{temp}
\end{table}%

\section{Discussion}\label{discussion}

The theoretical arguments  and the analysis of the LHC data presented above point to an  unconventional mechanism of apparent thermalization in high-energy collisions. The effective temperature $T_{th}$ deduced from the data has been found here to be non-universal and proportional to the hard scale of the collision $T$, i.e. to the momentum transfer, with $T \simeq 4.2\  T_{th}$. Strikingly, this conclusion seems to apply even to the Higgs boson production, suggesting that even in this very hard process the QCD radiation may be affected by thermalization. Moreover, we have found that the thermal component of the spectrum is entirely absent in diffractive production (even though many hadrons are still produced in this case) -- this again points to the non-universal, process-dependent, nature of thermalization. 
\vskip0.3cm

All of these features of the data seem to be consistent with the picture of thermalization induced by quantum entanglement. Indeed, in this scenario the effective temperature is proportional to the momentum transfer $Q$ in the collision that provides the UV cutoff for the quantum modes. This expectation agrees with our analysis of the inclusive charged hadron and Higgs boson transverse momentum distributions, in which the typical momentum transfers are vastly different. We have found that the thermal component is present in both cases, but the values of the effective temperature differ by over an order of magnitude\footnote{We stress once again that we do not imply that the Higgs boson is produced thermally, but rather that its transverse momentum distribution is affected by thermal radiation.}.
In diffractive production, one studies the coherent response of the entire proton, and there is no associated entanglement entropy \cite{kharzeev2017deep}. In this case, in the "thermalization through entanglement" picture advocated here, we expect to find no thermal component at all. This prediction is confirmed by the data on diffractive Drell-Yan production analyzed in this paper, as well as by the diffractive deep-inelastic scattering data  \cite{Bylinkin-Rostovtsev:2014}. 
\vskip0.3cm

These findings suggest a deep connection between quantum entanglement and thermalization in high-energy hadron collisions that has to be investigated further. On the experimental side, our study can be extended in several directions. In deep inelastic scattering at the future Electron Ion Collider, it would be necessary to combine the measurements of the structure functions with the study of hadronic final states, especially in the target fragmentation region. In proton-proton, proton-nucleus and nucleus-nucleus collisions at RHIC and LHC one can study the thermal component and the corresponding effective temperature in hard processes characterized by different momentum transfer. It would also be very interesting to investigate the dependence of the apparent thermalization on rapidity -- the picture presented in Fig. \ref{fig:sketch} suggests that thermalization is achieved faster if we perform a measurement in a smaller rapidity interval. 
\vskip0.3cm

It is clear that we are still very far from understanding thermalization in high-energy QCD, and much remains to be done both in theory and in experiment. Nevertheless, basing on the arguments and analysis presented above we believe that  ``thermalization through entanglement" emerges as a promising research direction that has to be pursued further.



\section{Acknowledgements}
The authors acknowledge support from the US Department of Energy,
Award E00176 (OKB) and  Contracts No. DE-FG- 88ER40388 and
DE-AC02-98CH10886 (DEK).

\section{References}
\bibliographystyle{unsrt}

\end{document}